\documentstyle[aps,prl,multicol,epsf]{revtex}
\begin{document}
\draft
\title{Dephasing Times in a Non-degenerate Two-Dimensional Electron Gas}
\author{I. Karakurt, D. Herman, H. Mathur, and A.J. Dahm}
\address{Department of Physics, Case Western Reserve University,
Cleveland, OH 44106-7079} 
%\email{ajd3@cwru.edu}

%97-01428.}
%\subjclass{}%
%\keywords{}%

\date{today}
\maketitle
%\dedicatory{}%
%\commby{}%
% ----------------------------------------------------------------
\begin{abstract}
Studies of weak localization by scattering from vapor atoms for
electrons on a liquid helium surface are reported.  There are
three contributions to the dephasing time.  Dephasing by the
motion of vapor atoms perpendicular to the surface is studied by
varying the holding field to change the characteristic width of
the electron layer at the surface.  A change in vapor density
alters the quasi-elastic scattering length and the contribution
to dephasing due to the motion of atoms both perpendicular and parallel 
to the surface.  Dephasing due to the electron-electron interaction is 
dependent on the electron density.
\end{abstract}
\pacs{PACS: 73.20.Fz, 73.20.Jc, 73.20.Dx }
\begin{multicols}{2}
\input epsf
% ----------------------------------------------------------------

Weak localization of degenerate electrons by elastic scattering
from static impurities has been a topic of serious study for the
last two decades \cite{Bergmann}. Recently there has been a revival
of interest in the damping of weak-localization in these systems
by the electron-electron interaction \cite{Mohanty,Golubev,Aleiner}. 
In comparison there have been few studies of weak localization 
in non-degenerate systems in which localization is due to 
quasi-elastic scattering from slowly moving impurities 
\cite{D'yakonov,Adams1,Adams2}. Adams and Paalanen explored 
both weak and strong localization of electrons on a solid hydrogen 
surface \cite{Adams1,Adams2,Adams3}.
Localization occurred as a result of scattering from surface
imperfections and from helium atoms that were introduced above the
surface. In our system electrons are confined to two dimensions above
a liquid helium surface. Weak-localization results from quasi-elastic
scattering from slowly moving helium vapor atoms. This system is 
particularly interesting because it possesses an unusual mechanism 
for damping quantum effects, namely the motion of the vapor atoms. 
At the same time there are a number of experimentally tunable
parameters (the electron density, the holding field that helps
confine the electrons to the helium surface, and the vapor density)
that can be varied to separate various damping mechanisms.
In this Letter we report a systematic investigation of the
dephasing times in this non-degenerate two-dimensional electron
gas.

Weak-localization is a quantum effect that results from 
constructive interference between closed electron paths and 
their time reversed counterparts. This constructive interference
increases the probability of back-scattering and results in an
increase in resistivity over the classical Drude value. 
In addition to the electron-electron interaction, in our system
weak-localization is damped by the slow motion of the 
helium vapor atoms.
The velocity of thermal helium atoms is 1\% of the electron 
velocity and the fractional change in
electron energy in a collision is $ \approx 10^{-2}$. 
There is an important distinction in the way in which
weak localization is suppressed by the vertical and
horizontal motions of the helium vapor atoms. Horizontal
motion changes the lengths of the paths introducing a 
random relative phase between a path and its approximate
time-reversed counterpart, thereby washing out their
interference \cite{Afonin1,Stephen1}. In contrast 
vertical motion suppresses weak localization because roughly 
speaking the scattering atom may not be present for both the forward
and return path, thereby reducing the weight of the interference
contribution. Below we estimate the dephasing
rate due to vertical motion of the vapor atoms; more details
will be given elsewhere \cite{Herman}. Similar ideas have been
expressed in Ref. \cite{Stephen2}, but the precise formula
we obtain is different. The corresponding discussion of 
horizontal motion is given in Refs. \cite{Afonin1,Stephen1}.

We use a Corbino geometry consisting of four electrodes located
beneath the helium liquid. The resistivity is measured by capacitively
coupling a low frequency ac current through the electron 
layer \cite{Mehrotra}. A normal field is applied to the 
inner three electrodes which are used for the resistivity
measurement. When the holding field $E_{\perp}$ is greater
than the saturated field $E_{s} = n e/2 \varepsilon_{0}$,
a voltage positive $V_{0}$ is applied to the outermost guard electrode
to compensate for fringing fields. We adjust the guard
voltage to maximize the signal. The signal amplitude decreases
if either the area of the third electrode covered with electrons
is reduced (which reduces the capacitance between the electrons
and the electrode) or if the diamater of the electron pool 
becomes sufficiently large that it capacitively couples to
the guard ring. Numerical calculation with this optimum 
value of $V_{0}$ indicates a nearly uniform electron density 
above the three inner electrodes.

For a non-degenerate, two-dimensional electron gas, the
longitudinal conductivity in a magnetic field B perpendicular to
the plane of electrons is given by \cite{Adams2,Altshuler}
\begin{eqnarray}\label{eq1}
\sigma_{xx} & = & \frac{-n_{0}e^{2}}{m(k_{B}T)^{2}}
                  \int_{E_{C}}^{\infty} d E
    \frac{e^{-E/k_{B}T}}
         {1+(\mu B)^{2}}\{
    E\tau_{0}
\nonumber \\
 & & -\frac{\hbar}{2\pi}[
    \Psi(\frac{\hbar m}{4eBE\tau_{0}^{2}}+\frac{1}{2})
   -  \Psi(\frac{\hbar
         m}{4eBE\tau_{0}\tau_{\phi}}+\frac{1}{2})]\}.
\end{eqnarray}
Here $n$ is the electron density, $\mu$ is the mobility, $\Psi$ is
the digamma function, $\tau_{0}$ is the quasi-elastic scattering
time, $\tau_{\phi}$ is the dephasing time, $E$ is the energy, and
$E_{c}$ is the

\epsfxsize=3.0in \epsfbox{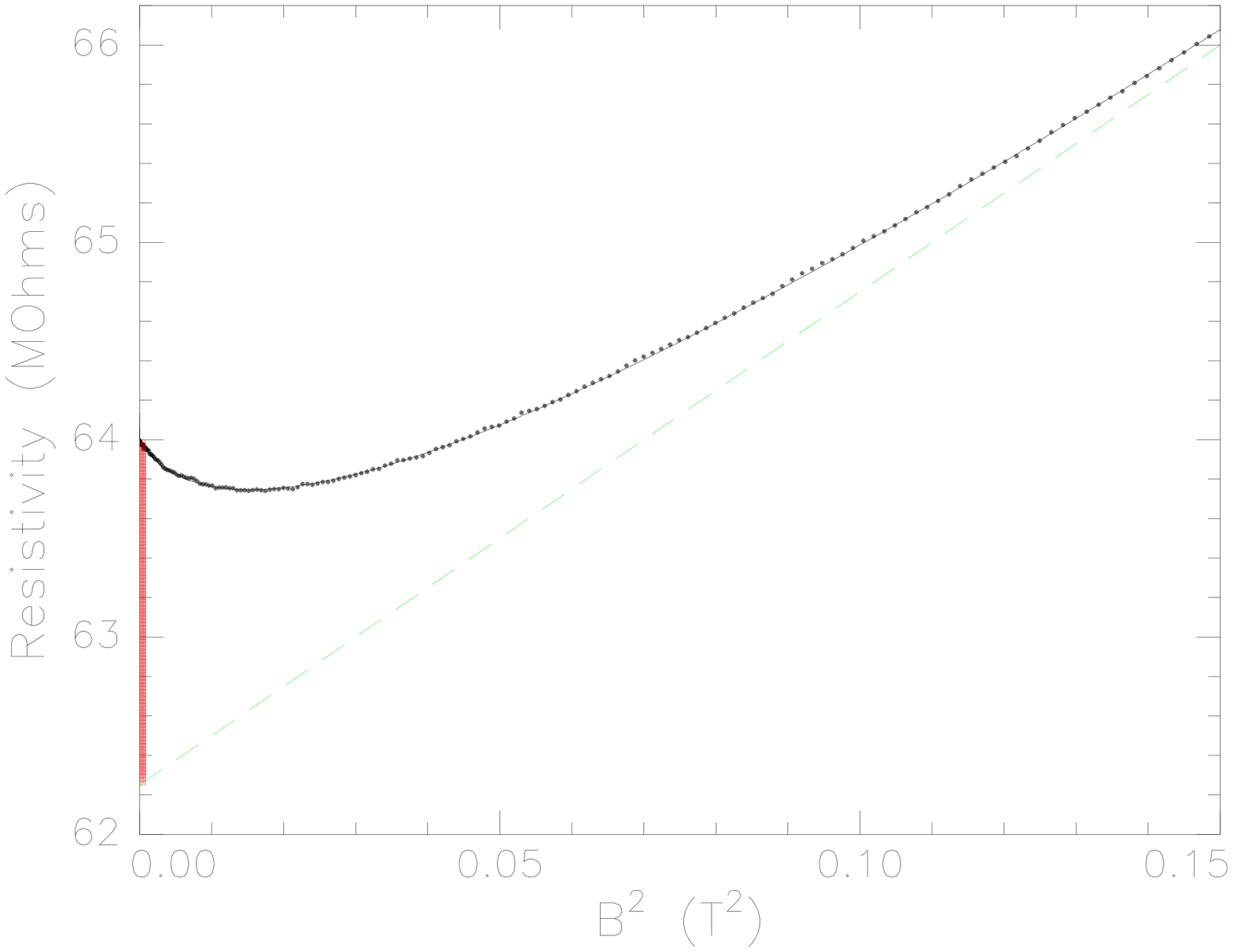}
\figure{Figure 1. 
$\rho_{xx}$ versus $B^{2}$; n = $1.7\times10^{11} m^{-2}$, T =
$2.15$ K, $\mu = 0.7 m^{2}$/Vs. The dashed line is the Drude
theory.} 

\vspace{2mm}

\noindent
cutoff energy below which electrons are localized.
The first term in the curly brackets gives the Drude resistivity,
and the second term gives the weak-localization correction. We 
assume that the total dephasing rate is given by
\begin{equation}
\tau_{\phi}^{-1} = \tau_{ee}^{-1} + \tau_{v}^{-1} + \tau_{h}^{-1}.
\end{equation}
Here $ \tau_{ee}$, $ \tau_{v}$ and $\tau_{h}$ are the dephasing times
due to the electron-electron interaction, vertical motion, and horizontal
motion, respectively.

The dephasing time $\tau_{\phi}$ is measured by fitting the
longitudnal magnetoresistance. Figure 1 shows a  
graph of the longitudinal resistivity $(1/\sigma_{xx})$ versus
magnetic field. The solid line is given by Eq.
(1) with the parameters $\tau_{0}$ and $\tau_{\phi}$ adjusted to
give the best fit \cite{gamma}. The dashed line is the Drude resistivity. In 
our analysis we calculate $E_{c}$ self-consistently from
the expression \cite{Adams1,Stephen1}
\begin{equation}\label{eq7}
E_{c} = (\hbar/2\pi\tau_{0})\ln(\tau_{\phi}/\tau_{0}).
\end{equation}
The fits are relatively insensitive to the value of $E_{c}$
since $E_{c} \leq 250 mK \ll T$.

The dephasing due to the electron-electron interaction is caused by
the fluctuations in the electric field due to other electrons.
These fluctuations are controlled by thermally 
excited plasma oscillations. We therefore assume that the 
dephasing rate is inversely proportional to the characteristic
plasma frequency
\begin{equation}
\omega_{cp} =
\sqrt{n^{3/2}e^{2}/(2m\overline{\varepsilon})};\hspace{7mm}
\overline{\varepsilon} = (\varepsilon + \varepsilon_{0}))/2.
\end{equation}
To verify this assumption and to separate the contribution
of electron-electron interaction to $\tau_{\phi}$ we measured
the dephasing time as a function of electron density. These 
data are shown in Fig. 2 where $ \tau_{\phi}$ is plotted as
a function of plasma frequency. The data are fit by
\begin{equation}
\tau_{\phi} = \tau_{A}/(1 + \alpha\omega_{cp}\tau_{A}).
\end{equation}

\epsfxsize=3.0in \epsfbox{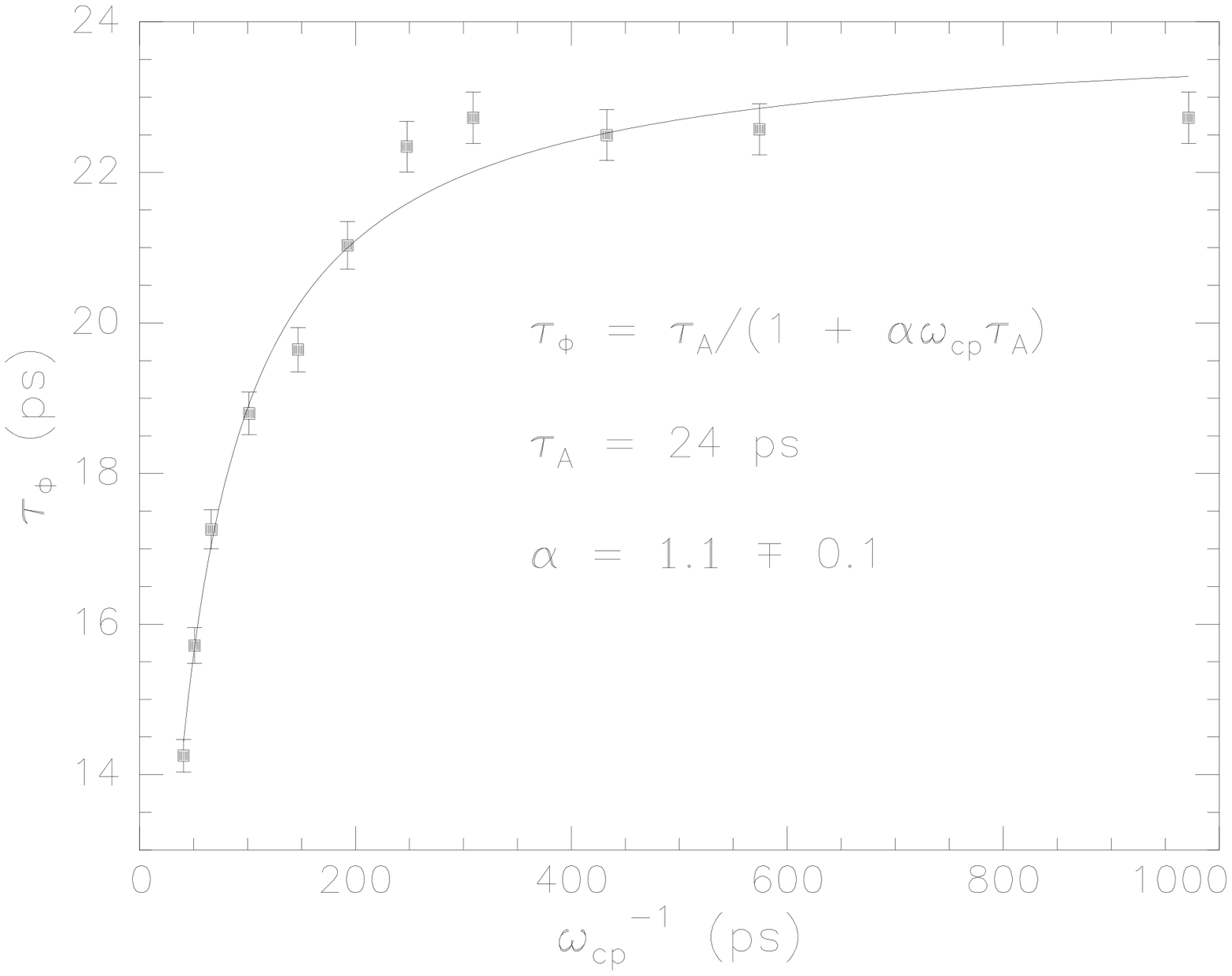}
\figure{Figure 2.Dephasing time versus inverse plasma frequency. 
T = $1.96$ K.}

\vspace{2mm}

\noindent
This equation follows from Eq. (2) with $\tau_{A}$ defined as
$\tau_{A}^{-1} \equiv \tau_{v}^{-1} + \tau_{h}^{-1}$ and
$\tau_{ee} = 1/\alpha\omega_{cp}$.  The best fit for the parameter
$\alpha$ is $1.1 \pm 0.1$.

Figure 2 shows that our data are consistent with the assumption
that the timescale for dephasing via electron-electron interaction
is set by the plasma frequency. Further support for this assumption
comes from measurements of the electronic velocity auto-correlation
time which is also found to be set by the inverse plasma frequency
\cite{Zipfel}. We hope this finding will stimulate the development
of a theory of dephasing by electron-electron interaction that is 
applicable to a non-degenerate electron gas.

We turn now to the damping due to the motion
of the vapor atoms. The theoretical expression for the
dephasing time due to horizontal motion is \cite{Afonin1,Stephen1}
\begin{equation}
\tau_{h} = (g \tau_0 \tau_{\lambda}^{2} )^{1/3}; 
\hspace{4mm} \tau_{\lambda} = \lambda/\sqrt{2 k_{B} T/m }.
\end{equation}
Here $\lambda$ is the de Broglie wavelength of the electron and
the theoretical value of $g = 6$. The analogous expression for
vertical motion is
\begin{equation}
\tau_{v} = (f \tau_0 \tau_{z}^{2})^{1/3};
\hspace{4mm} 
\tau_{z} = b/\sqrt{ k_{B} T / m }.
\end{equation} 
Here $b$ is a measure of the width of the vertical subband
wavefunction of the electrons and the theoretical value of
$ f = 9/2 $. It is a known function of the
holding electric field (see discussion following Eq. 9). Eq (7)
has not appeared before in the literature; we sketch its 
derivation below.

\epsfxsize=3.0in \epsfbox{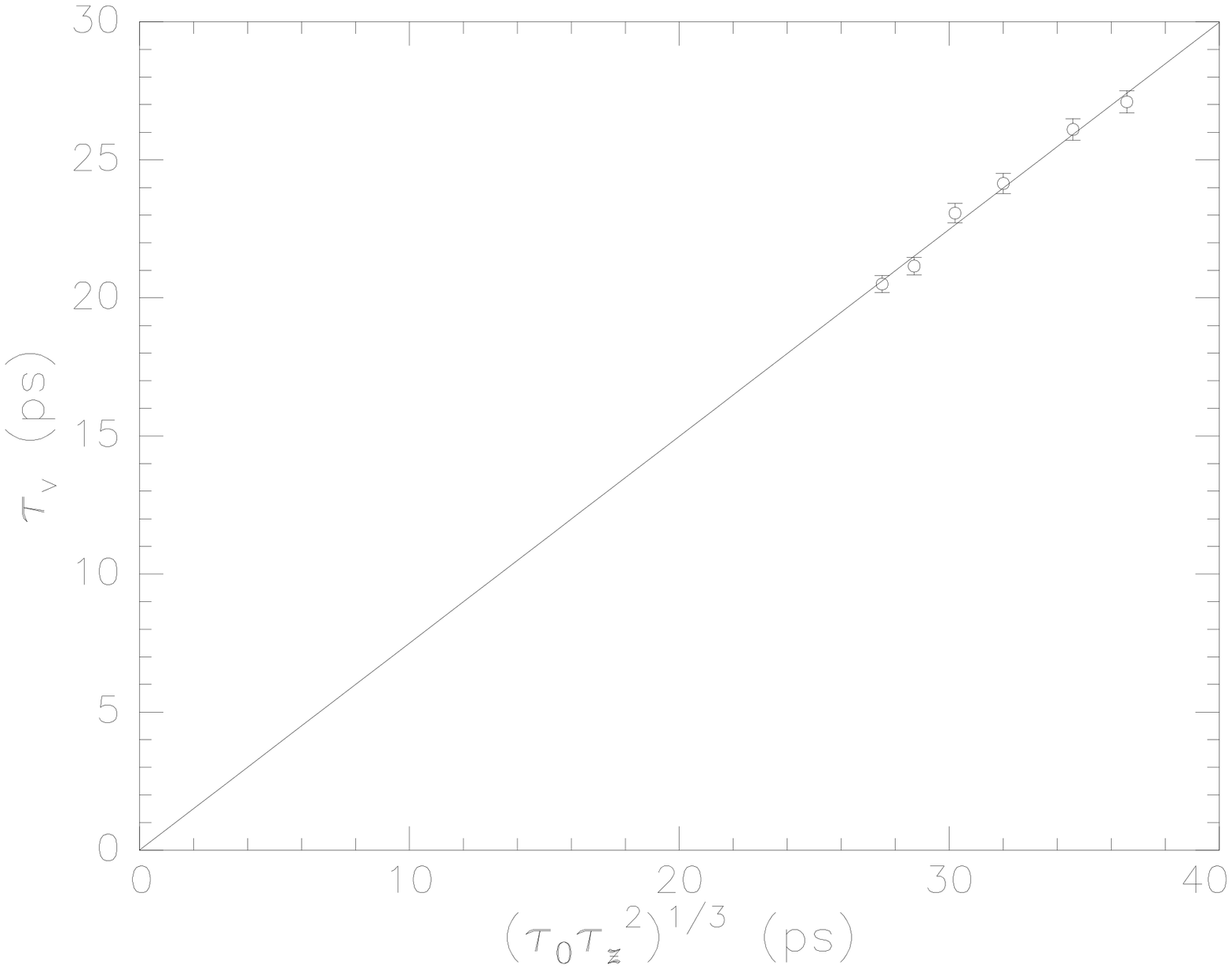}
\figure{Figure 3. 
$\tau_{v}$ versus $(\tau_{0}\tau_{z}^{2})^{1/3}$. T = $2.165$ K.
The solid line is a fit to theory with f = $0.4$, and g = $1.3$.}

\vspace{2mm}

To test the theoretical expressions, Eqs. (6-7), and to separate
$\tau_v$ and $\tau_h$ we vary the characteristic width $b$ of the 
electronic wave function by changing the holding field. 
The range of $b$ was limited by microphonic induced instabilities
of the charged surface at small $n$ and large $E_{\perp}$. 
We calculate
$\tau_{v}$ from Eq. (2) using the empirical value of $\tau_{ee}$ and
the theoretical value of $\tau_{h}$ (Eq. 6) but with $ g $ as an adjustable
parameter. The calculated values of $\tau_{v}$ are plotted as a
function of $(\tau_{0} \tau_{z}^{2})^{1/3}$ (see Fig. 3) and the
parameter g is adjusted until the best linear fit through the
data passes through the origin. This yields values of $f=0.4 \pm 0.1$ 
and $g=1.3 \pm 0.3$. For the measurements shown in Fig. 3
the product $k_{T} l$ was 1.6-2.4 with $k_{T} =
\sqrt{2 m K_{B} T }/\hbar$, and the product $\omega_{cp} \tau_0$
was in the range 0.05-0.08. Thus, dephasing 
is dominated by dephasing due to the motion
of helium atoms in these data. The measured dephasing times $ \tau_v $
and $\tau_h$ are comparable because $b$ and $\lambda$ are comparable
(at zero applied holding field $ b = 7.6 $nm while $\lambda = 14$nm
at 2.1 K.).

A second comparison to the theoretical expressions, Eqs.(6-7)
comes from measuring the dependence of $\tau_{\phi}$ on 
the electron-atom scattering time $\tau_{0}$ studied by 
changing the vapor density. Combining Eqs. (6) and (7) shows
\begin{equation}
\tau_{A}^{-1} = [ ( f \tau_{z}^{2} )^{-1/3} +
(g \tau_{\lambda}^{2} )^{-1/3} ] \tau_{0}^{-1/3}.
\end{equation}
The coefficient of $\tau_{0}^{-1/3}$ contains temperature
dependent parameters $\tau_{z}$ and $\tau_{\lambda}$. A graph
of $\tau_{A}^{-1}$ is shown in Fig 4. The values of $\tau_{A}^{-1}$
increase with increasing $\tau_0^{-1/3}$, but the fits to theory
are poor. Figure 4 shows a fit with $f = 0.8$ and $ g = 1.0$. Similar
fits can be obtained with $f$ and $g$ up to $f=0.4$ and $g=4$,
but the data cannot be fit to the values $f=0.4, g= 1.0$ obtained
from the fit to Fig 3. For these
measurements the product $k_{T} l$ was in the range 2.4-7.5.

\epsfxsize=3.0in \epsfbox{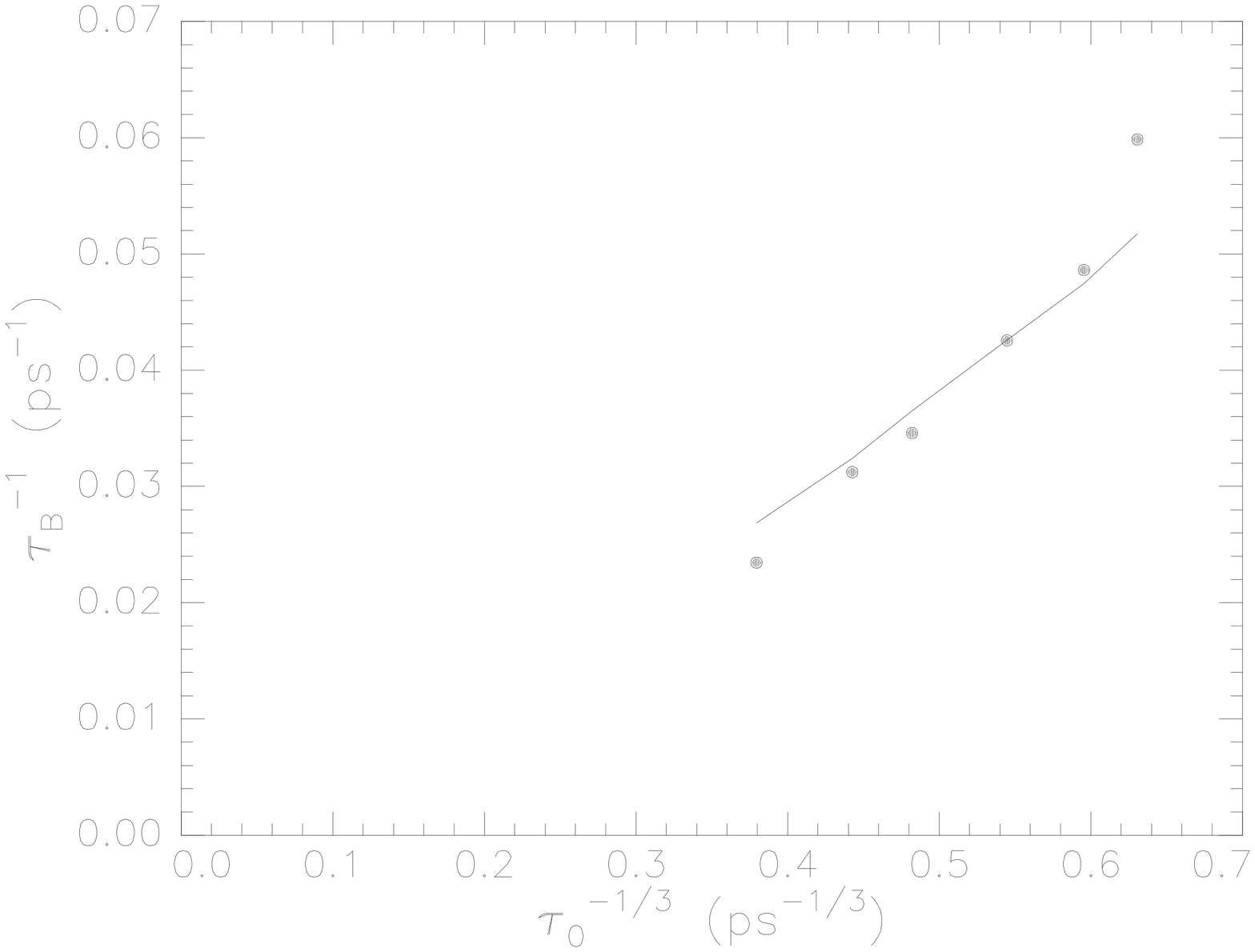}
\figure{ Figure 4. The
quantity $ \tau_{A}^{-1}$ versus $\tau_{0}^{-1/3}$.}

\vspace{2mm}

We also studied the variation of the scattering time $\tau_{0}
\propto \mu$ with the holding field $E_{\perp}$. Figure 5 
shows a plot of the mobility as a function of the width
$b$ for electrons on both isotopes of helium. Curves represent
scaled theoretical values. The data are inconsistent with
theory, which predicts the mobility to be linear in $b$
\cite{Saitoh}. Values of $k_{T} l$ were in the range
1.6-2.4 and 1.0-1.3 for $^4$He and $^3$He, respectively, and
$E_c$ was as large as 600mK for $^3$He. The differences
in behaviour for the two isotopes may be related to the
close approach to strong localization for $^3$He.

We turn now to the derivation of the formula for damping
due to vertical motion of the vapor atoms (Eq. 7; see Ref.
\cite{Herman} for more details).
If we treat the helium vapor atoms as hard-core potentials, the
contribution of a path to the return amplitude is a product of the
amplitude to scatter off the first atom, multiplied by the
amplitude to go to the second atom, multiplied by the amplitude to
scatter off the second atom, and so on around the loop.  Let
$A(z)$ be the amplitude to scatter from an atom at a height z
above the liquid helium surface.  We choose
\begin{eqnarray}\label{eq2}
A(z) & = & \frac{16 \pi a \hbar^2}{m b^3} 
z^2 \exp \left( - \frac{2 z}{b} \right) \hspace{3mm}
{\rm for} \hspace{2mm} z > 0 \nonumber \\
 & = & 0 \hspace{3mm} {\rm for} \hspace{2mm} z < 0.
\end{eqnarray}
This is derived by taking the vertical subband wavefunction of the
electrons to be of the variational form $ \psi(z) = 2
b^{-3/2} z \exp(-z/b)$ \cite{Schroedinger}.  The helium atoms are
treated as hard-core potentials; $a$ is the s-wave scattering
length, and the variational parameter $b$ is a function of the
applied field.

We now assume the helium atoms are allowed to move vertically
\cite{Nayak}.  Since a given atom is encountered at different
times on the forward and return paths we must consider
\begin{equation}\label{eq3}
Q(t) \equiv  \langle A(z) A(z + v t) \rangle.
\end{equation}
Here $t$ is the difference in the times at which the atom is

\epsfxsize=3.0in \epsfbox{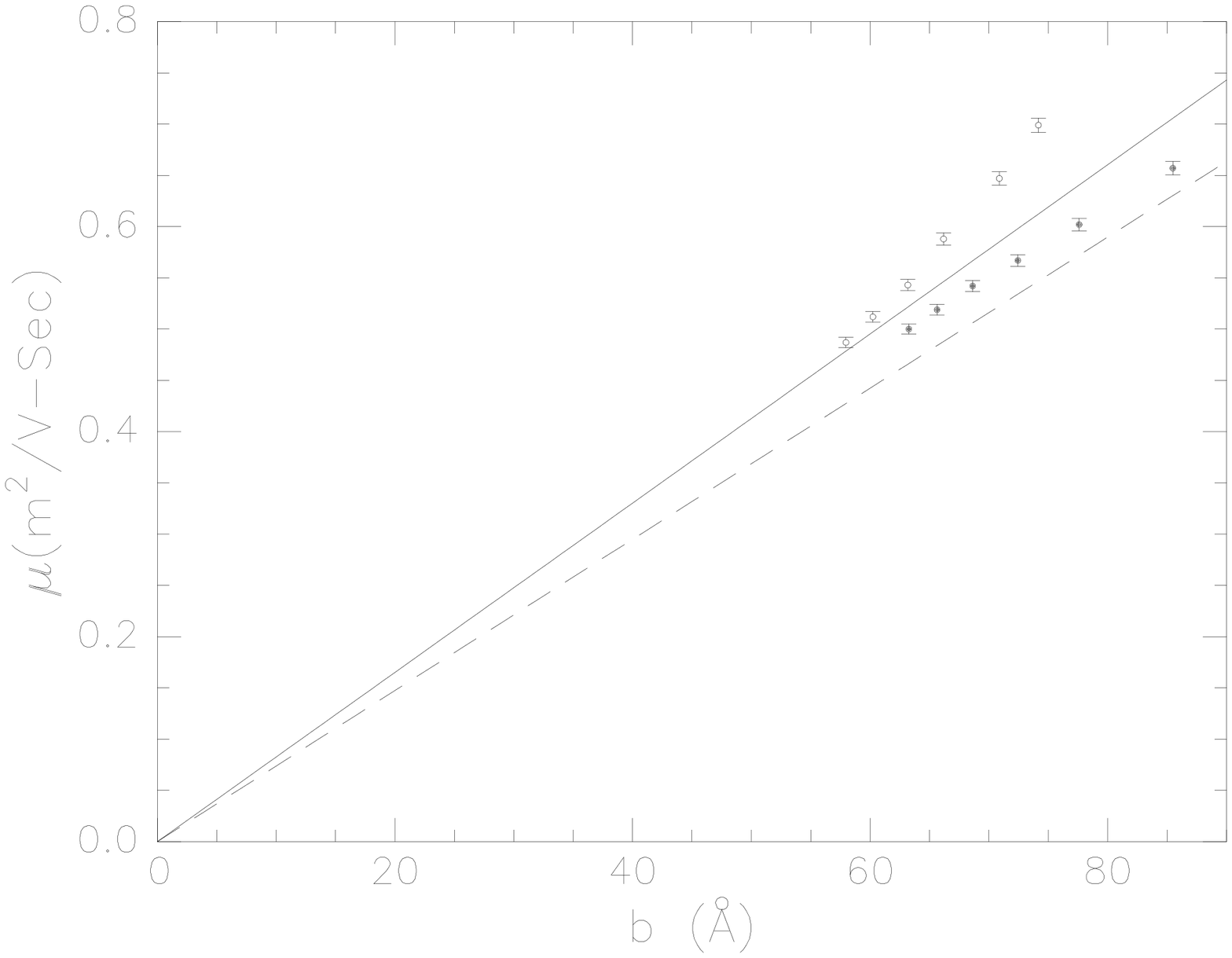}
\figure{ Figure 5. Mobility versus the variational parameter $b$. 
Open symbols, $^{4}$He at $T$ = 2.165K; closed symbols, $^{3}$He at
$T$ = 1.26K . Curves are theory/$1.75$:
solid - $^{4}$He; dashed - $^{3}$He.}

\noindent
encountered on the forward and return paths.  The atom is assumed
to move ballistically at vertical speed $v$ for this time.
$\langle \ldots \rangle$ denotes an average over all possible
configurations of the helium atom (vertical position is assumed to
be uniformly distributed and vertical speed is given by the
Maxwell-Boltzmann formula).  The interference between the forward
and return path is then reduced by the factor $ q(t) \equiv
Q(t)/Q(0) $ due to the motion of this atom.  A path of duration
$t$ encounters $ t/\tau_0 $ atoms; hence its interference with its
time reversed partner is reduced by a factor $ q(t)^{t/\tau_0}$
due to the vertical motion of all the atoms.  This estimate is
improved by noting that the difference in times at which an atom
is encountered by the forward and return paths is not the same for
all atoms: it varies from zero (for atoms in the middle of the
path) to $t$ (for atoms at the ends).

Using Eqs. (9) and (10) we find that the contribution of paths of
duration $t$ is reduced by $\exp(-t^3/\tau_{v}^3)$ \cite{form} and
$\tau_{v}^{z}$ is given by Eq. (7). In general, damping factors vary as
$\exp(- C t^{\gamma})$.  For electron-electron interactions, $\gamma
= 1$; for both horizontal and vertical motion of the helium atoms,
$\gamma = 3$. 

In conclusion, we have succeeded in separating the three
contributions to the dephasing times.  The times $\tau_{v}$ and
$\tau_{h}$ are found to be consistent with the predicted
functional forms. The experimentally determined values of
the numerical coefficients are $ f = 0.4 - 0.8 $ and
$ g = 1.0 - 1.5 $ (values based on the analysis of Fig
3 which we believe provides a more reliable estimate than
Fig 4). These values are an order of magnitude smaller than
the corresponding theoretical values. 
The reduction in $\tau_{v}$ with an increase in holding field 
has a simple explanation.  Increasing
the field reduces the width of the volume occupied by electrons
and, therefore, enhances the escape of atoms from this volume.
Nothing was known regarding the dephasing time due to electric
field fluctuations of other electrons.  We found $\tau_{ee}$
empirically to be $\approx\omega_{cp}^{-1}$, the only obvious
characteristic time associated with the electron gas.  

We thank V.I. Gurevich and M.J. Stephen for helpful
discussions. This work was supported in part by NSF Grants
DMR 97-01428 (I.K. and A.J.D.) and DMR 98-04983 (D.H. and H.M.),
and by the Alfred P. Sloan Foundation (H.M.).

% ----------------------------------------------------------------
%\bibliographystyle{amsplain}
%\bibliography{References}

\begin{references}

\bibitem{Bergmann}See G. Bergmann, Phys. Rep. {\bf 107}, 1 (1984);
P.A. Lee and T.V. Ramakrishnan, Rev. Mod. Phys. {\bf 57}, 287
(1985); S. Chakravarty and A. Schmid, Phys. Rep. {\bf 140}, 193
(1989).
\bibitem{Mohanty}P. Mohanty, E.M.Q. Jariwala, and R.A. Webb, Phys.
Rev. Lett. {\bf 78}, 3366 (1997); P. Mohanty and R.A. Webb, Phys.
Rev. B {\bf 55}, 13452 (1997).
\bibitem{Golubev} D.S. Golubev and A.D. Zaikin, Phys. Rev. Lett.
{\bf 81}, 1074 (1998); {\bf 82}, 3191 (1999); Phys. Rev. B {\bf
59}, 9195 (1999).
\bibitem{Aleiner}I. L. Aleiner, B.L. Altschuler, and M.E. Gershenson,
Phys. Rev. Lett. {\bf 82}, 3190 (1999); cond-mat/9808053.
\bibitem{D'yakonov}A.M. D'yakonov and Ya.V. Kopelevich, JETP
{\bf 47}, 259 (1988).
\bibitem{Adams1}P. W. Adams and M.A. Paalanen, Phys. Rev. Lett.
{\bf 58}, 2106 (1987).
\bibitem{Adams2}P. W. Adams and M.A. Paalanen, Phys. Rev. B {\bf 39},
4733 (1989).
\bibitem{Adams3}P.W. Adams, Phys. Rev. Lett. {\bf 65}, 3333 (1990).
\bibitem{Afonin1}V.V. Afonin, Yu.M. Gal'perin, V.I. Gurevich, and
A. Schmid, Phys. Rev. A {\bf 36}, 5729 (1987).
\bibitem{Stephen1}M.J. Stephen, PRB {\bf 36}, 5663 (1987).
\bibitem{Herman}D. Herman and H. Mathur, unpublished.
\bibitem{Stephen2}M.J. Stephen, Phys. Rev. B {\bf 40}, 2600 (1989).
\bibitem{Mehrotra}R. Mehrotra and A.J. Dahm, J. Low Temp. Phys.
{\bf67}, 115 (1987).  In Sec. $2.2$ of this paper the definitions
should be $\zeta_{1} = z_{1}$ and $\zeta_{2} = z_{2}$.  In Eqs.
$(16) - (20)$, $I$ should be replaced by $I' = I/2\pi$, and the
expressions for $Y$ in Eqs. $(24) - (27)$ should be multipled by
$2\pi$.
\bibitem{Altshuler} B.L. Altshuler, D. Khmel'nitzkii, A.I. Larkin,
and P.A. Lee, Phys. Rev. B {\bf 22}, 5142 (1980).
\bibitem{gamma} Eq (1) for the magnetoresistance is strictly
applicable only for $ \gamma = 1$ (see discussion following
Eq 10 for a definition of $ \gamma $). Use of the correct
expression for $ \gamma = 3 $ \cite{Altshuler,Afonin2} 
would lead to small shifts in
the measured values of $\tau_{\phi}$.
The expression for $\gamma = 3$ is very cumbersome to apply.
Other workers \cite{Adams2} have found Eq. (1) gives a 
satisfactory fit even when the damping is dominated by
vertical motion of helium atoms.
\bibitem{Afonin2}V.V. Afonin, Yu.M. Gal'perin, and V.I. Gurevich,
Phys. Rev. B {\bf 33}, 8841 (1986).
\bibitem{Zipfel}C.L. Zipfel, T.R. Brown, and C.C. Grimes, Phys. Rev.
Lett. {\bf 37}, 1760 (1976).
\bibitem{Saitoh} M. Saitoh , J. Phys. Soc. Jpn. {\bf 42}, 201 (1977).
\bibitem{Schroedinger}We have solved Schr\"{o}dinger's equation
numerically for an electron on helium in the presence of a holding
field.  For the holding fields used, $E_{\perp} \leq 580 V/cm$,
the variational wavefunction gives the average value of the z
co-ordinate and the full width at half maximum of
$\mid\psi\mid^{2}$ to within $1\%$.  However, the maximum of the
wavefunction shifts by as much as $11\%$ compared to the numerical
result.
\bibitem{Nayak} We assume that vapor atoms that strike the liquid
surface do not reflect since the sticking probability is known to
be $\approx 1$.  V.U. Nayak, N. Masuhara, and D.O.
Edwards, Phys. Rev. Lett. {\bf 50}, 990 (1983).
\bibitem{form}This functional form is valid when
$\tau_{z} \gg \tau_{0}$, a condition that is also necessary for
weak localization.

\end{references}

\end{multicols}

\end{document}